\documentclass[fleqn,usenatbib]{mnras}

\usepackage{newtxtext,newtxmath}
\usepackage{multirow}
\usepackage{multicol}
\usepackage[T1]{fontenc}

\DeclareRobustCommand{\VAN}[3]{#2}
\let\VANthebibliography\thebibliography
\def\thebibliography{\DeclareRobustCommand{\VAN}[3]{##3}\VANthebibliography}

\usepackage{graphicx}	% Including figure files
\usepackage{amsmath}	% Advanced maths commands

\title[Rapid neutron star cooling triggered by dark matter]{Rapid neutron star cooling triggered by dark matter}

\author[A. Ávila et al.]{
Afonso Ávila,$^{1}$\thanks{afonso.avila@student.uc.pt}
Edoardo Giangrandi,$^{1,2}$\thanks{edoardo.giangrandi@uni-potsdam.de}
Violetta Sagun$^{1}$\thanks{violetta.sagun@uc.pt}
Oleksii Ivanytskyi$^{3}$\thanks{oleksii.ivanytskyi@uwr.edu.pl}
and Constança Providência$^{1}$\thanks{cp@uc.pt}
\\
$^{1}$CFisUC, Department of Physics, University of Coimbra, Rua Larga P-3004-516, Coimbra, Portugal\\
$^{2}$Institut für Physik und Astronomie, Universität Potsdam, Karl-Liebknecht-Str.24-25, Potsdam, Germany\\
$^{3}$Incubator of Scientific Excellence---Centre for Simulations of Superdense
Fluids, University of Wrocław, 50-204, Wroclaw, Poland\\
}

\pubyear{2024}

\begin{document}
\label{firstpage}
\pagerange{\pageref{firstpage}--\pageref{lastpage}}
\maketitle

\begin{abstract}
We study the effect of asymmetric fermionic dark matter (DM) on the thermal evolution of neutron stars (NSs). No interaction between DM and baryonic matter is assumed, except the gravitational one. Using the two-fluid formalism, we show that DM accumulated in the core of a star pulls inwards the outer baryonic layers of the star, increasing the baryonic density in the NS core. As a result, it significantly affects the star's thermal evolution by triggering an early onset of the direct Urca process and modifying the photon emission from the surface caused by the decrease of the radius. Thus, due to the gravitational pull of DM, 
the direct Urca process becomes kinematically allowed for stars with lower masses. Based on these results, we discuss the importance of NS observations at different distances from the Galactic center. Since the DM distribution peaks towards the Galactic center, NSs in this region are expected to contain higher DM fractions that could lead to a different cooling behavior. 
\end{abstract}


\section{Introduction}
\label{sec:Introduction}

Extremely high gravitational field and compactness inside neutron stars (NSs) make them a perfect laboratory to study the strongly interacting matter, test General Relativity and physics beyond the Standard Model~\citep{Baym:2017whm,Kramer:2021jcw}. Throughout their entire lifetime, stars could accumulate a sizeable amount of dark matter (DM) in their interior, which will impact the matter distribution, masses, radii, evolution, etc.~\citep{Scott:2008uw,Lopes:2019jca}. At the end of its evolution, a main sequence star of 8-20 M$_\odot$ undergoes a supernova explosion, creating an NS~\citep{Heger_2003}. The former is from the gravitational collapse of molecular cloud regions, which exceed the Jeans limit. The proto-cloud may already present traces of DM, facilitating the collapse and giving rise to newly born stars with a sizeable amount of DM~\citep{Yang:2020bkh}. Once the star is born, DM particles could be further accreted from a surrounding medium, leading to an even higher DM fraction inside the object~\citep{Brito_2015,Kouvaris:2010jy}. A steady DM accretion requires a long timescale and a high DM fraction in the surrounding medium. The maximum amount of the total accreted mass of DM is 0.01\% of the star's total mass in the most central part of the Galaxy~\citep{Ivanytskyi:2019wxd}. On the other hand, rapid accumulation and a significant increase of the DM fraction inside the star could occur while passing through an extremely dense region of primordial DM clumps in a subhalo~\citep{Bramante:2021dyx}. As predicted by many cosmological models, primordial density perturbations could result in a large fraction of DM forming gravitationally collapsed objects residing in subhalos~\citep{Erickcek:2011us,Buckley:2017ttd}.

At the end of the stellar evolution, the star eventually reaches the iron-core stage, undergoing a core-collapse supernova explosion. During this incredibly energetic event, DM might be created and further accrued inside the remnant, i.e. an NS~\citep{Meyer:2020vzy}.
Once DM is trapped in the gravitational field of an NS, it may lead to different configurations depending on the DM properties: a core or halo configuration. In the former scenario, DM forms a compact core in the inner regions of an NS. A stronger gravitational pull by the inner core leads to more compact and denser configurations, characterized by smaller maximum gravitational masses and radii compared to the purely baryonic star. Thus, these configurations may be seen as an apparent softening of the baryonic equation of state (EoS)~\citep{Giangrandi:2022wht}. NSs with a compact DM core are also harder to deform, an effect that can be tested by future gravitational wave (GW) detections via the tidal polarizability $\Lambda$~\citep{Giangrandi:2022wht,Sagun:2021oml,Karkevandi:2021ygv,Tolos_2021,Diedrichs:2023trk,Thakur:2024mxs}. It could manifest itself by a supplementary peak or strong oscillation mode in the post-merger GW spectrum~\citep{Ellis:2017jgp, Bezares:2019jcb}, production of an exotic waveform~\citep{Giudice:2016zpa}, or modification of the kilonova ejecta~\citep{Emma:2022xjs}. The latter probes are expected to be possible with the next generation of GW detectors, i.e., the Einstein Telescope (ET)~\citep{Punturo:2010zz}, Cosmic Explorer (CE)~\citep{Mills:2017urp}, and NEMO~\citep{Ackley:2020atn}. On the other hand, when the radius of the DM component exceeds the baryonic one, a halo structure is formed, fully embedding the star. This leads to an increase in the star's gravitational mass, mimicking a stiffening of the baryonic EoS~\citep{Sagun:2022ezx}. The halo is easier to be deformed due to the diluted DM distribution surrounding the star~\citep{Ivanytskyi:2019wxd, Shakeri:2022dwg}, affecting the tidal polarizability $\Lambda$ of the star~\citep{Sagun:2021oml,Diedrichs:2023trk}.

Another way to probe the presence of DM in NSs is by studying their thermal evolution. Standard NS cooling\footnote{By the standard thermal evolution we mean cooling by the emission of Standard Model particles, i.e. neutrinos and photons.} occurs via a combination of thermal radiation from the surface and neutrino emission from the interior, making it possible to test the particle composition and properties of matter through x-ray observations. In fact, the thermal evolution of compact stars contains very rich and complicated phenomena at different stages. Thus, it can be divided into three stages: newly born NS with the thermally decoupled core and crust (age $\lesssim100$ yr)~\citep{Sales:2020aad}, neutrino emission dominant stage (age $100-10^{6}$ yr) and the photon emission dominant stage (age $\gtrsim 10^{6}$ yr)~\citep{Page:2004fy}. 

The first stage corresponds to the time required for the core and crust of the newly born NS to become thermally equilibrated, the thermal relaxation time. A different composition of the NS's core and crust results in a substantial variance in their thermal conductivity and specific heat~\citep{Lattimer:1994glx}. Neutrinos, emitted from the core, create a cold front that advances toward the surface. When it reaches the surface, its temperature suddenly drops, marking the start of a core-crust thermal connection. Prior to this, the surface temperature of the star remains unchanged as neutrinos slowly diffuse from the core towards the surface, supplying energy that counterbalances cooling.

Further on, NS cooling is mainly defined by the particle composition of the NS core. Particularly, the amount of protons (or, equivalently, electrons and muons together) defines whether the star undergoes a rapid cooling governed by the direct Urca (DU) process, or cools down slowly/intermediately via the nucleon-nucleon bremsstrahlung, modified Urca (MU), and Cooper pair breaking and formation (PBF) processes~\citep{Page:2005fq}. In stars where the DU process takes place, a significant temperature drop is observed, heavily modifying their thermal evolution. The underlying EoS determines whether the DU process of neutron $\beta$-decay and its inverse process can occur in the NS interior~\citep{Lattimer:1991ib,Page:2005fq, Potekhin:2015qsa}. For these reactions to take place, the Fermi momenta of the involved particles have to satisfy the kinematic restriction of the triangle inequality, $p_{Fp}+p_{Fe}\ge p_{Fn}$ given in terms of the Fermi momenta of protons, electrons and neutrons (for n,p,e matter). This condition ensures that for strongly degenerate fermions the reaction is constrained by the Pauli blocking principle, meaning that it can only take place when the energies of the particles involved are close to their Fermi energies. By considering charge neutrality and the relation between the Fermi momenta and the number density of each particle, the proton fraction, $Y_{p}$, should be above $\sim 11\%$~\citep{Lattimer:1991ib}. When the threshold condition is not satisfied, the less effective MU process is the dominant neutrino emission process due to the presence of a spectator nucleon that mediates the reaction.

When the star is cool enough so that neutrino emission becomes less relevant, the photon emission from its surface becomes the dominant cooling process. Typically, the star continues to cool down until it becomes invisible to X-ray telescopes and the peak of the black body radiation shifts towards longer wavelengths. However, additional contributions that go beyond the minimal cooling paradigm could alter the above-mentioned scenario by contributing with additional heating or cooling channel. While the accretion of matter from a companion star~\citep{Wijnands:2017jsc} and magnetic field decay~\citep{Aguilera:2007dy} only deposit energy into the star, the presence of DM, depending on the considered candidate, could contribute to either cooling or heating. Through emission-evaporation of light DM from the NS core and/or surface, DM could carry away energy further cooling the star~\citep{Kumar:2022amh}. While the emission of DM from the star is mostly considered for light particles that can freely escape, e.g. axions, heavy DM (with mass above the MeV scale) could also evaporate from the star's surface~\citep{Garani:2021feo}. Many studies have been conducted to model the effect of axion emission on NS and proto-NS~\citep{Dietrich:2019shr} thermal evolution. Axions produced within the NS cores in, e.g., nucleon bremsstrahlung or PBF processes, leave the star, contributing to its cooling ~\citep{Sedrakian:2015krq,Buschmann:2019pfp, Buschmann:2021juv}. 

On the other hand, DM may heat the star during the accretion or self-annihilating, depositing energy in the system~\citep{Baym_2018,Motta_2018b,Motta_2018,Berryman:2022zic}. DM via scattering with the Standard Model particles may deposit the kinetic energy gained falling into a steep NS gravitational potential, the so-called ``dark kinetic heating''~\citep{Baryakhtar:2017dbj, Raj:2017wrv}. Moreover, trapped symmetric DM could annihilate in the NS interior~\citep{2010PhRvD..81l3521D}, or decay, leading to a further heating channel. Depending on the model, this process can be observed in either middle-time~\citep{AngelesPerez-Garcia:2022qzs} or late-time heating~\citep{Hamaguchi:2019oev}. As shown by~\citet{Kouvaris_2008} the NS cooling curves at the late stage may have a plateau corresponding to DM self-annihilation. This scenario appears to be the easiest to test, as the effect of nuclear superfluidity/superconductivity and magnetic field are negligible at this age. The major contribution to the cooling of old NSs comes from the photon emission, while the neutrino cooling stage sensitive to a particle composition is suppressed. Therefore, a possible heating mechanism of NSs due to DM self-annihilation could be probed by the increasing statistics on observational data of old NSs, in the range from soft x-ray to infrared bands, including the operating James Webb Space Telescope~\citep{Chatterjee:2022dhp} and the forthcoming Thirty Meter Telescope (TMT) and Extremely Large Telescope (ELT)~\citep{TMTInternationalScienceDevelopmentTeamsTMTScienceAdvisoryCommittee:2015pvw}. 

Alternatively, asymmetric DM considered in this work interacts with BM only through gravity, and, therefore, does not contribute to either neutrino, photon emission, or self-annihilation. It is assumed that DM is accrued during the previous stages of a star's evolution and remains unchanged during the NS lifetime. Accounting for the accretion of DM during the NS equilibrated stage with consequent kinetic heating requires a more extensive study that we leave for future work.

The DM fraction up to a few percent could be accrued during the previous stages of star evolution by several mechanisms, including the 
DM production during a supernova explosion, the rapid accumulation in the main-sequence star and proto-NS while passing the extremely dense subhalos~\citep{Bramante:2021dyx} or be present in the star progenitor due to the primordial clump of DM formed during the DM-dominated era~\citep{Yang:2020bkh}.

The paper is organized as follows. In Section \ref{sec:Models}, we present models for the BM and DM components. In Section \ref{sec:CoolingThings}, we discuss the main processes ruling the NS thermal evolution and how we implemented the second fluid in our calculations. In Section \ref{sec:Results} the main results are presented, including the DM effects on the DU threshold and the cooling curves. Section \ref{sec:Conclusion} includes conclusions and discussions of the smoking gun signal of the presence of DM that could be tested in the near future.
Throughout the article, we utilize the unit system in which $\hbar=c=G=1$.

\begin{table*}
\begin{tabular}{c|llllllllll}
\hline	
& ~~~$n_{0}$ & ~$E/A$ &~~ $K_0$  & ~ $E_{sym}$ & ~~$L$  & $M_{max}$ & $R_{1.4}$&$M_{DU}$ &$n_{DU}$ &$Y_p^{DU}$\\
&  [fm$^{-3}$]  & [MeV] & [MeV]  & [MeV] & [MeV] & [M$_{\odot}$] & [km]& [M$_\odot$] &[fm$^{-3}$]  &\\
\hline
IST & 0.16  & -16.00  & 201.0 & 30.0  & 93.19& 2.084 & 11.4 &1.908& 0.869 &0.11 \\
\hline
FSU2R& 0.1505 & -16.28  & 238.0 & 30.7 & 46.9 & 2.048 & 12.8 &1.921 &0.608&0.14\\
\hline
\end{tabular}
\caption{Parameters of the IST and FSU2R models. The table includes the saturation density $n_{0}$, energy per baryon $E/A$, incompressibility factor $K_{0}$, symmetry energy $E_{sym}$, and its slope at the saturation density $L$, as well as the maximum gravitational mass $M_{max}$, radius of the 1.4 M$_{\odot}$ star, and the NS mass, baryonic density and proton fraction that characterize the onset of the DU process.}
\label{tab1}
\end{table*} 

\section{DM-admixed stars}
\label{sec:Models}

In this section, we first summarize the properties of the two BM models chosen, IST and FSU2R, and the DM model. We next review the TOV equations for two-fluid configuration and show the mass-radius curves for the BM models chosen with different fractions of DM.

\subsection{BM models}
\label{subsec:BarMatter}

To address the uncertainties of the BM EoS we chose two models with different 
nuclear matter properties at the saturation density. Particularly, with the different values of the symmetry energy slope L and incompressibility factor $K_{0}=9(\frac{\partial p~}{\partial n_{BM}})_{n_0}$ at saturation.

The first model is based on the induced surface tension (IST) approach, formulated with an explicit account of the hard-core repulsion among the particles. It was shown by~\citet{Sagun:2013moa} that in a dense medium, a short-range repulsive interaction among the particles induces an additional contribution to the single-particle energy, the IST term. At high densities, the IST contribution is negligibly small compared to other terms in the single-particle energy, and the excluded volume treatment of hard-core repulsion is switched to the proper volume regime. In the dilute gas limit, this approach recovers the first four virial coefficients of hard spheres. In the IST EoS the hard-core radius of nucleons was obtained
from the fit of the heavy-ion collision data~\citep{Sagun:2017eye, Bugaev:2021pwn}. Application of the IST EoS to the nuclear liquid-gas phase transition with its critical endpoint allowed constraining the parameters of the eigensurface tension of nucleons~\citep{Sagun:2016nlv}, while the attraction and symmetry energy terms were fitted to the NS observables~\citep{Sagun:2018cpi}. Hereby, we use the Set B of the IST EoS developed in~\cite{NSOscillationsEoS}. 

The second considered model is the nucleonic relativistic mean-field FSU2R EoS~\citep{Tolos:2017lgv}, which is a further development of the FSU2 approach \citep{Chen:2014sca} with a softer symmetry energy and neutron matter pressure. This allows the FSU2R EoS to describe NS with radii smaller than in the case of FSU2 EoS. The parameters of this model were fitted to the binding energies, charge radii, and monopole response of atomic nuclei across the periodic table. It equally well reproduces the properties of nuclear matter and finite nuclei.

Both considered models, the IST and FSU2R, reproduce the flow constraint~\citep{Danielewicz:2002pu,Ivanytskyi:2017pkt}, tidal deformability of GW170817 and the observational data from the GW190425 binary NS merger~\citep{LIGOScientific:2018cki,LIGOScientific:2020aai}, NICER mass-radius constraints~\citep{Miller:2019cac,Riley:2019yda,Raaijmakers:2019dks,Miller:2021qha,Riley:2021pdl} and the heaviest pulsars measurements. Table.~\ref{tab1} summarizes the main model parameters.

For the realistic description of the outer layers, the IST and FSU2R EoSs are supplemented by the Haensel-Zdunik (HZ) EoS for the outer crust and the Negele-Vautherin (NV) EoS for the inner crust~\citep{1990A&A...227..431H,Negele:1971vb}.

\subsection{DM model}
\label{subsec:DM}

DM is modeled as a relativistic Fermi gas of non-interacting particles with spin one-half, which has been extensively studied in the literature, e.g.~\cite{Nelson:2018xtr,Ivanytskyi:2019wxd,Sagun:2021oml}. The expressions for the pressure and energy density in the grand canonical ensemble can be written as
\begin{equation}
\begin{cases}
    p_{DM} = \frac{g_{DM}}{48\pi^2}[\mu_{DM} k_{DM} (2\mu_{DM}^2-5m_{DM}^2)+3m_{DM}^4\ln(\frac{\mu_{DM}+k_{DM}}{m_{DM}})],\\
    \varepsilon_{DM}=\mu_{DM} n_{DM} -p_{DM},
\end{cases}
\end{equation}
where $g_{DM}=2$, $\mu_{DM}$, $n_{DM}=\partial p_{DM} / \partial \mu_{DM}$ and $k_{DM}=\sqrt{\mu_{DM}^2-m_{DM}^2}\theta(\mu_{DM}-m_{DM})$ are the DM degeneracy factor, chemical potential, number density, and Fermi momentum, respectively. We consider DM to be at zero temperature. It is well motivated considering that DM is accrued long enough to thermalize with the NS matter and reach the same vanishing temperature.

\subsection{Two-fluid configurations}
\label{subsec:TOV}

Since DM interacts with BM only gravitationally, the stress-energy tensors of the two fluids are conserved separately, which allows us to write down the two coupled Tolman-Oppenheimer-Volkoff (TOV) equations~\citep{PhysRev.55.364,PhysRev.55.374}
\begin{equation}\label{TOV}
\frac{dp_i}{dr}=-\frac{(\varepsilon_i +p_i)(M_\mathrm{tot}+4\pi r^3p_\mathrm{tot})}{r^2\left(1-{2M_\mathrm{tot}}/{r}\right)},
\end{equation}
where the subscript index $i=\mathrm{BM},\mathrm{DM}$ labels the components, $M_\mathrm{tot}=M_{DM}+M_{BM}$ with $dM_i/dr=4\pi r^2\varepsilon_i$ is the total gravitational mass enclosed to the sphere of radius $r$ and $p_\mathrm{tot}=p_{DM}+p_{BM}$ is the total pressure. 

\begin{figure}
    \centering
    \includegraphics[width=\columnwidth]{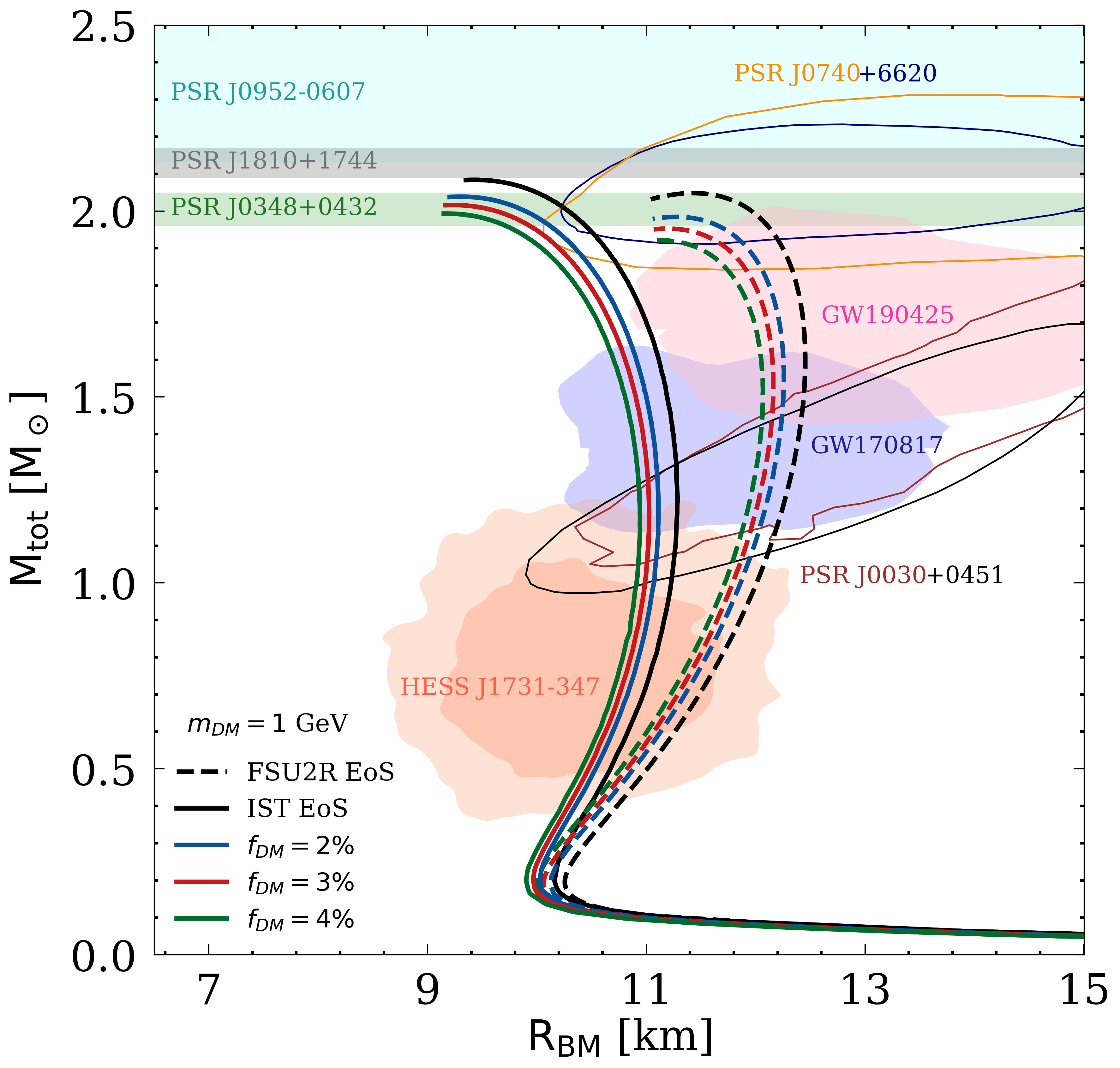}
    \caption{Total gravitational mass of the DM-admixed NS as a function of its baryonic radius $R_{BM}$ calculated for the DM particle mass $m_{DM}$=1 GeV. Solid and dashed black curves correspond to pure BM stars described by the IST EoS and FSU2R EoS, respectively.  Royal blue, red, and green curves characterize relative DM fractions equal to 2\%, 3\%, and 4\%, respectively. Green, gray, and cyan bands represent $1\sigma$ constraints on mass of PSR J0348+0432~\citep{Antoniadis:2013pzd}, PSR J1810+1744~\citep{Romani:2021xmb}, and PSR J0952-0607~\citep{Romani:2022jhd}. Red and black contours show the NICER measurements of PSR J0030+0451~\citep{Miller:2019cac, Riley:2019yda}, while yellow and blue contours correspond to the PSR J0740+6620 measurement~\citep{Raaijmakers:2021uju, Miller:2021qha}. LIGO-Virgo collaboration observations of GW170817~\citep{LIGOScientific:2018cki} and GW190425~\citep{LIGOScientific:2020aai} binary NS mergers are shown in blue and pink. The $1\sigma$ and $2\sigma$ contours of HESS J1731-347~\citep{Doroshenko2022} are plotted in dark and light orange.}
    \label{fig:MRCurveIST}
\end{figure}

\begin{figure}
    \centering
\includegraphics[width=\columnwidth]{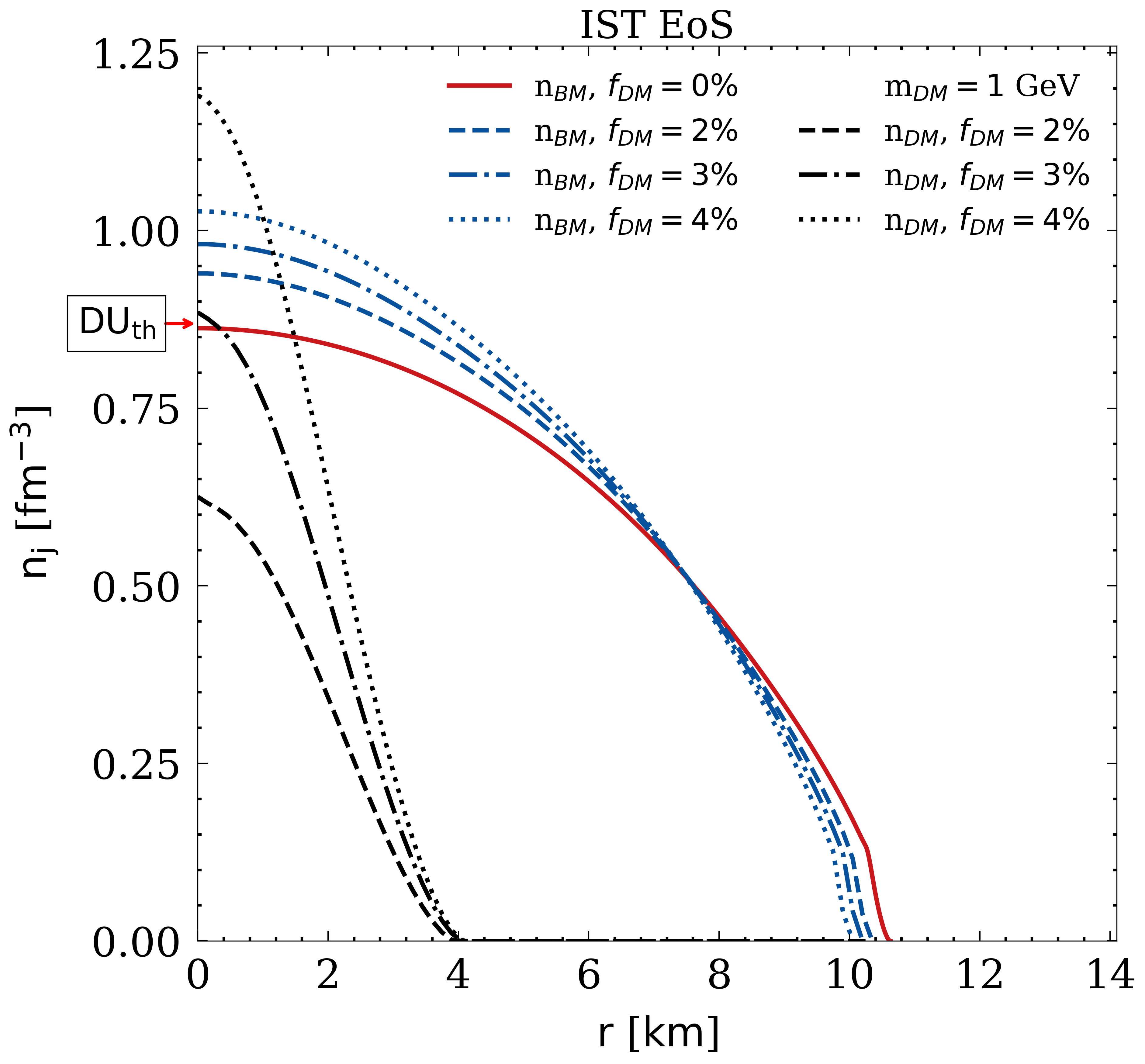}\\
\includegraphics[width=\columnwidth]{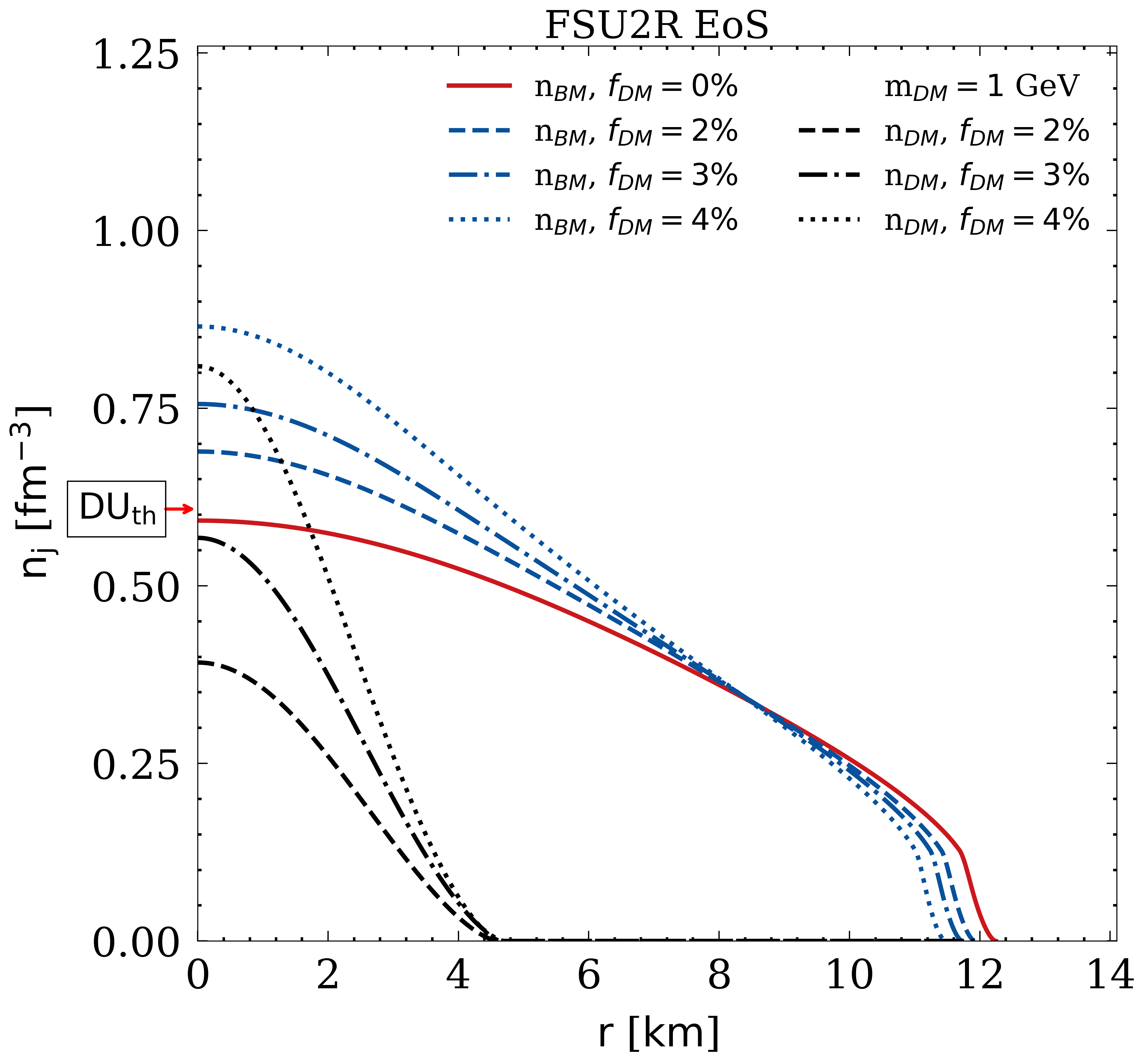}
    \caption{The split number density profiles for BM and DM for the $1.9\ M_\odot$ star. The impact of DM of different fractions on the star's profile is shown for the IST EoS (upper panel) and FSU2R EoS (bottom panel). The calculations are performed for DM particle mass $m_{DM}=1$ GeV. The red arrow on the y-axis indicates the density of the DU threshold.}
    \label{fig:ProfilesISTFSU2RDM}
\end{figure}

By fixing the values of the central chemical potentials of each of the components we are able to obtain NSs of different mass admixed with a different amount of DM. It is convenient to work in the grand canonical ensemble, as the chemical potentials of two components are related to each other, as 
\begin{equation}
    \frac{d \ln \mu_{BM}}{dr}=\frac{d \ln \mu_{DM}}{dr} = -\frac{M_\mathrm{tot}+4\pi r^3 p_\mathrm{tot}}{r^2(1-2M_\mathrm{tot}/r)}.
\end{equation}
The full derivation of this relation is presented in ~\cite{Ivanytskyi:2019wxd}. It shows, that for a given stellar configuration the chemical potentials of two components scale proportionally, i.e. $\mu_{BM}\propto\mu_{DM}$, which simplifies solving the system of two coupled TOV equations.
After integration of the TOV equations, the gravitational masses of each of the components are recovered, and it is possible to define the DM mass fraction as $f_{DM} = \frac{M_{DM}}{M_\mathrm{tot}}$. 

The mass-radius relations computed for the IST EoS (solid curves) and FSU2R EoS (dashed curves) are shown in Figure~\ref{fig:MRCurveIST}. The effect of increasing the DM faction from 2\% to 4\% is shown in different colors. As is seen, the presence of the DM core reduces the gravitational maximum mass of NSs and their radii, being similar to an apparent softening of the BM EoS. Also included in the figure are the constraints obtained from different NS observations as identified in the caption. 

For our study, we chose a single value of the DM particle mass, $m_{DM}$=1 GeV, that leads to the formation of the distinctive DM core. However, for completeness, the scans over the relative DM fraction and particle's mass are performed in Section~\ref{subsec:Results_scan}. In Fig.~\ref{fig:ProfilesISTFSU2RDM} the depicted number densities for BM and DM as a function of radius for the IST EoS (upper panel) and FSU2R EoS (bottom panel) show the relation between the compactness of the DM core and BM redistribution. Thus, with an increase of the DM fraction, a more compact DM core pulls BM inward, leading to a smaller star radius. This effect will be further discussed in the context of the photon luminosity in Section~\ref{subsec:Results_cooling}. 
The radii of the 1.9 M$_\odot$ DM-admixed stars with the respective profiles shown in Fig.~\ref{fig:ProfilesISTFSU2RDM} are given in Table~\ref{Table:radii}. 

\section{NS Cooling}
\label{sec:CoolingThings}

The evolution of NSs is governed by the thermal balance equation~\citep{Page:2005fq} 
\begin{equation}
C_{\mathrm{v}} \frac{d T{_s^\infty}}{d t}=-L_\nu^\infty-L_\gamma^\infty \pm H^\infty,
\label{Eq:4}
\end{equation}
where $C_\mathrm{v}$ is the total specific heat of the stellar matter, $T{_s^\infty}$, $L_\nu^\infty$, $L_\gamma^\infty$ are the redshifted surface temperature, neutrino luminosity, and photon luminosity, respectively. 
The last term in Eq.~(\ref{Eq:4}) accounts for any additional source of heat (sign ``$+$'') or carrying energy away (sign ``$-$'') discussed in Introduction. In this study, no additional source of heating or cooling is considered, therefore, $H^\infty\equiv 0$.

The photon luminosity depends on the star's radius as $L_\gamma = 4\pi \mathrm{R^2} \sigma T{_s}^4$. The redshifted functions are obtained by multiplication by the $e^{\Phi}$ factor, being the $g_{tt}$ component of the Schwarzschild metric. It includes the metric function $\Phi$, which dependence on the radial coordinate $r$, and can be obtained by solving the following ODE
\begin{equation}
\frac{d\Phi}{dr}= -\frac{dp_{\mathrm{tot}}}{dr}\frac{1}{\varepsilon_\mathrm{tot}+p_\mathrm{tot}},
\label{Eq:5}
\end{equation}
where $p_\mathrm{tot}=p_{BM}+p_{DM}$, $\varepsilon_\mathrm{tot}=\varepsilon_{BM}+\varepsilon_{DM}$ are the total pressure, and energy density, respectively.

In comparison to the photon luminosity that follows the typical black body radiation law, the neutrino emissivity in each process depends on factors such as density, temperature, and the Cooper pairing between nucleons. 
When the DU process is not allowed, the MU process makes a dominant contribution in removing heat in the form of $\nu$ emission, unless $n$ and $p$ are in a paired state. The latter substantially suppresses the rates of neutrino emission. Thus, neutrons/protons in the NS interior exist in superfluid/superconducting states by forming Cooper pairs. Despite the suppression of the rates of neutrino emission, the breaking of Cooper pairs is a new source of $\nu$ balancing the former one. As a result, Cooper pairs are constantly breaking and forming (the PBF process) providing the medium cooling rate~\citep{Potekhin:2015qsa}. This process gets activated for $n$ and $p$ when the temperature reaches their respective critical values. The pairing strength is defined by the gap parameter, its form, and the peak of the PBF emissivity, varying for the adopted gap model. Free neutrons in the inner crust and protons in the core pair in a singlet-state ($^1S_0$), while neutrons in the core are expected to undergo a triplet-state pairing ($^3P_2$)~\citep{Bardeen:1957mv,Page:2005fq}. Consequently, the application of various models of nucleon pairing to explore the cooling of NSs also offers the opportunity to gain deeper insights into the properties of NS matter.

{\bf Implementation.} We use the publicly available thermal evolution code \texttt{NScool}~\citep{Neutronstarcoolingcode}.
To account for the gravitational impact of DM, profiles were generated for each target mass using the two-fluid formalism. As asymmetric DM does not directly contribute to a star's cooling, all particle species remain unchanged. Hence, there are two types of variables used by the one-fluid framework of \texttt{NSCool}: the ones related to the BM EoS that do not include the DM contribution, i.e. baryon density and particle fractions, and the ones that include such a term, i.e., total pressure, total energy density, total gravitational mass and the metric functions as the two fluids exist in the same spacetime.

{\bf Data.} The observational data of Cassiopeia A (Cas A) supernova remnant are depicted as source 0. The insets on the two panels of Fig.~\ref{fig:CassA} indicate the temperature measured using Chandra ACIS-S in GRADED and FAINT modes with 1$\sigma$ error bars. Following~\cite{Shternin:2022rti}, we show the data points obtained for varying and fixed hydrogen column density $\mathrm{N_H}$ during the observational time, leading to different $T{_s^\infty}$ values. The rest of the data shown in Figs.~\ref{fig:CoolingISTFSU2RDM}-\ref{fig:CassA} were taken from ~\cite{Beznogov:2014yia}.  We consider 2$\sigma$ error bars for the available data, otherwise a factor of 0.5 and 2 for both temperature and age, excluding the upper limits. The sources are: 1 - PSR J0205+6449 (in 3C58), 2 - PSR B0531+21 (Crab), 3 - PSR J1119-6127, 4 - RX J0822-4300 (in PupA), 5 - PSR J1357-6429, 6 - PSR B1706-44, 7 - PSR B0833-45 (Vela), 9 - PSR J0538+2817, 10 - PSR B2334+61, 11 - PSR B0656+14, 12 - PSR B0633+1748 (Geminga), 13 - PSR J1741-2054, 14 - RX J1856.4-3754, 15 - PSR J0357+3205 (Morla), 16 - PSR B1055-52, 17 - PSR J2043+2740, 18 - RX J0720.4-3125. The surface temperature of the object 8 - XMMU J1731-347 from~\cite{Beznogov:2014yia} was substituted by the updated results HESS J1731-347 from~\citep{Doroshenko2022}. 

\section{Results}
\label{sec:Results}

\subsection{The DU onset}
\label{subsec:Results_DU}

\begin{figure}
    \centering
    \includegraphics[width=\columnwidth]{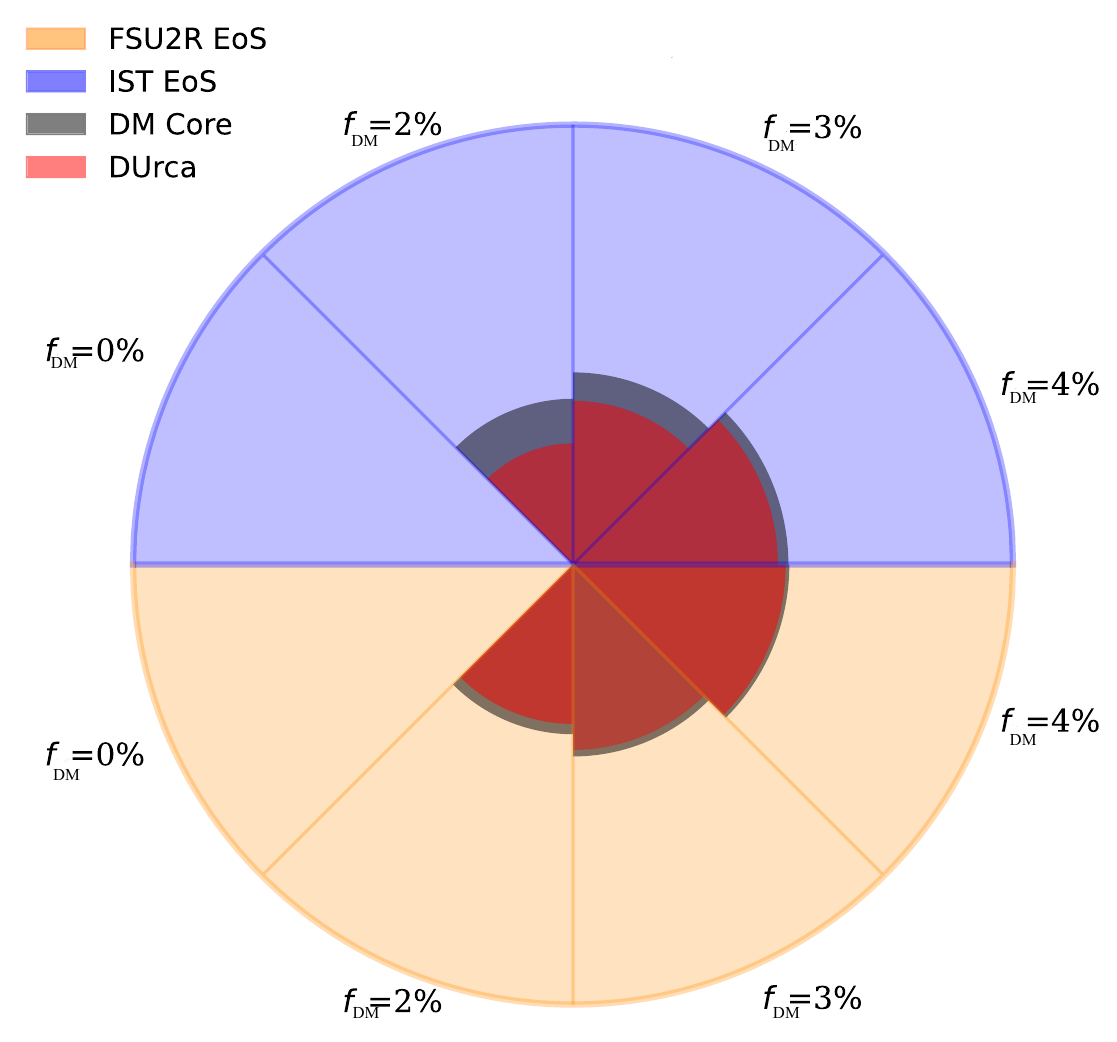}
    \caption{Stellar configurations with different DM fractions for the IST EoS (upper part) and FSU2R EoS (lower part). The size of the DU region and DM core are depicted in dark grey and light red, respectively. For comparison, the radii are normalized to the outermost baryonic radius of each configuration and are given in Table~\ref{Table:radii}. All configurations correspond to NSs with a total gravitational mass of 1.9 M$_\odot$.
    }
    \label{fig:AmericanPlot}
\end{figure}

We start with a pure BM 1.9 M$_\odot$ NS, where DU is still not active in its center, and add DM particles of mass $m_{DM}$=1 GeV. As shown schematically in Fig.~\ref{fig:AmericanPlot}, the accrued DM inside the core triggers the DU process. An increase of the DM fraction causes an extension of the region where the neutron $\beta$-decay and its inverse process are allowed (these two regions are depicted in Fig.~\ref{fig:AmericanPlot} in dark grey and light red, respectively). The rest of the star matter outside of the light red region cools down via the slow/medium processes. For a better comparison, in Fig.~\ref{fig:AmericanPlot} all the radii are normalized to the outermost baryonic radius of the considered configuration. The physical values of the radii and total gravitational masses of the considered configurations in this figure and Figs.~\ref{fig:CoolingISTFSU2RDM}-\ref{fig:CassA} are presented in Table~\ref{Table:radii}.

\begin{table}[t]
\resizebox{0.7\columnwidth}{!}{%
\begin{tabular}{|c|cccc|}
\hline
\multirow{2}{*}{IST EoS}   & \multicolumn{4}{c|}{$\mathrm{f_{DM}}$}                                                                            \\ \cline{2-5} 
                           & \multicolumn{1}{c|}{$0\%$} & \multicolumn{1}{c|}{$2\%$} & \multicolumn{1}{c|}{$3\%$} & $4\%$                      \\ \hline
$M_\mathrm{tot}$ [$\mathrm{M_\odot}$]     & \multicolumn{4}{c|}{$R_\mathrm{BM}$ [km]}                                                                                       \\ \hline
1.20                       & \multicolumn{1}{c|}{11.29} & \multicolumn{1}{c|}{11.11} & \multicolumn{1}{c|}{11.03} & 10.94                      \\ \hline
1.60                       & \multicolumn{1}{c|}{11.10} & \multicolumn{1}{c|}{10.91} & \multicolumn{1}{c|}{10.81} & 10.70                      \\ \hline
1.90                       & \multicolumn{1}{c|}{10.58} & \multicolumn{1}{c|}{10.35} & \multicolumn{1}{c|}{10.20} & 10.05                      \\ \hline
\multirow{2}{*}{FSU2R EoS} & \multicolumn{4}{c|}{$\mathrm{f_{DM}}$}                                                                            \\ \cline{2-5} 
                           & \multicolumn{1}{c|}{$0\%$} & \multicolumn{1}{c|}{$2\%$} & \multicolumn{1}{c|}{$3\%$} & $4\%$                      \\ \hline
$M_\mathrm{tot}$ [$\mathrm{M_\odot}$]     & \multicolumn{4}{c|}{$R_\mathrm{BM}$ [km]}                                                                                       \\ \hline
1.20                       & \multicolumn{1}{c|}{12.18} & \multicolumn{1}{c|}{12.09} & \multicolumn{1}{c|}{12.01} & 11.93                      \\ \hline
1.60                       & \multicolumn{1}{c|}{12.39} & \multicolumn{1}{c|}{12.25} & \multicolumn{1}{c|}{12.15} & 12.05                      \\ \hline
1.90                       & \multicolumn{1}{c|}{12.16} & \multicolumn{1}{c|}{11.93} & \multicolumn{1}{c|}{11.73} & \multicolumn{1}{l|}{11.47} \\ \hline
\end{tabular}%
}
\caption{Parameters of the considered DM-admixed stars in Figs. \ref{fig:CoolingISTFSU2RDM} and \ref{fig:CassA}.
}
\label{Table:radii}
\end{table}

\begin{figure}
    \centering
    \includegraphics[width=\columnwidth]{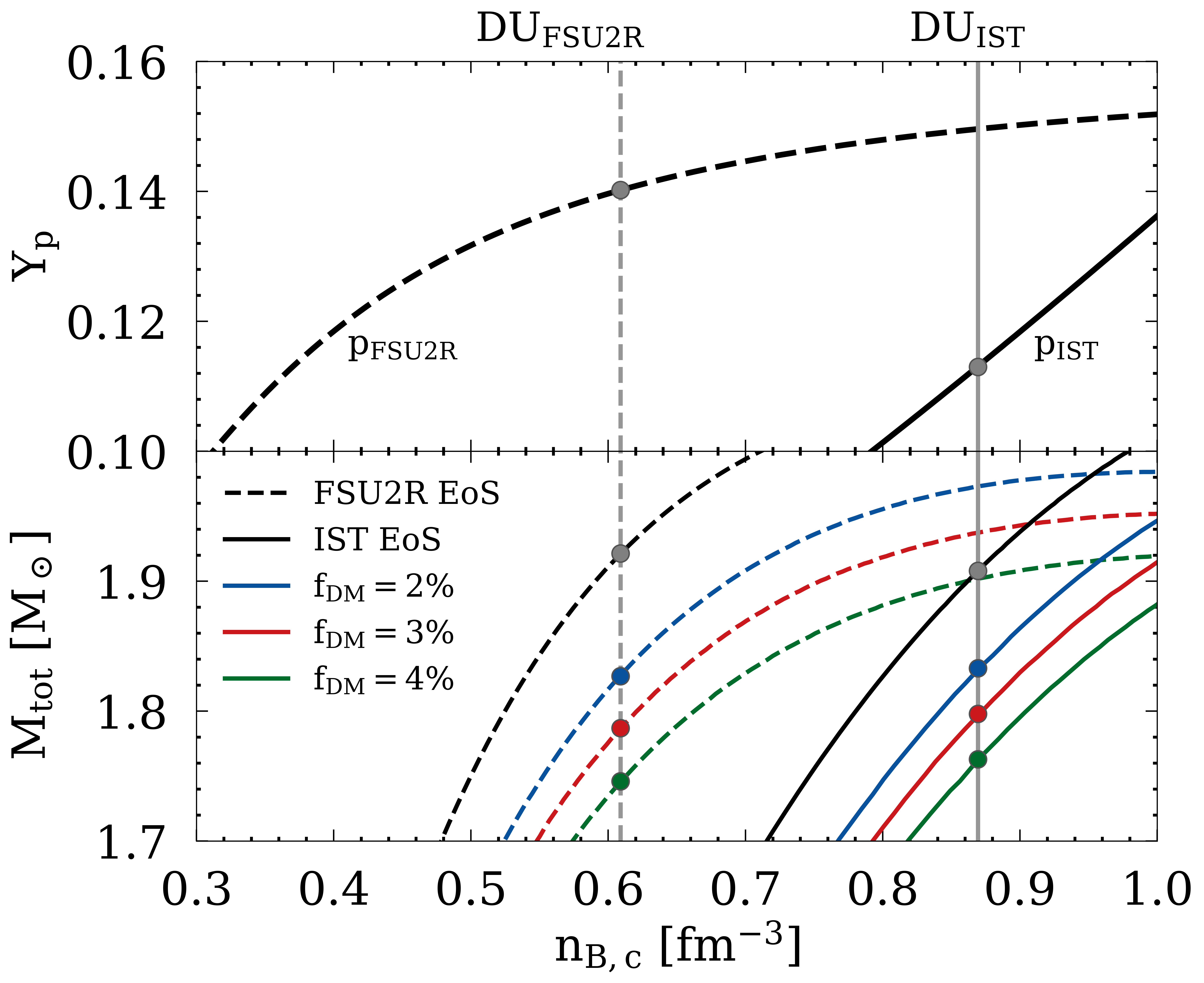}
    \caption{The proton fraction, $Y_{p}$, and total gravitational mass of NSs as a function of the central baryonic density, $n_{B,c}$. The curves are equivalent to those in Fig.~\ref{fig:MRCurveIST}. The vertical solid (for the IST EoS) and dashed (for the FSU2R EoS) grey lines correspond to the central BM densities of the stars at which the DU process is activated. The intersection points correspond to the minimal total gravitational mass at which the nucleonic DU process threshold is satisfied.}
    \label{fig:DUrcaOnset}
\end{figure}

The asymmetric DM interacting only via gravity with the Standard Model particles does not directly affect the neutrino or photon emission. Therefore, the DU process is kinematically allowed at the same proton fraction and central baryonic density for stars with and without DM. As it can be seen in Fig.~\ref{fig:DUrcaOnset} the 
presence of DM alters the value of the total gravitational mass at which the DU process is kinematically allowed. With an increase of the DM fraction, we see a drastic reduction in the total gravitational mass for both models. The total gravitational mass of the star with the triggered DU gets dramatically reduced. Thus, for the FSU2R EoS the mass of the star at the DU onset, $M_{DU}$, drops from 1.92 M$_{\odot}$ to  1.83 M$_{\odot}$, 1.79 M$_{\odot}$, and 1.75 M$_{\odot}$ for 2\%, 3\%, and 4\% of DM, respectively. For pure BM stars modeled by the IST EoS, the DU processes are allowed at 1.91M$_{\odot}$. The vertical solid (for the IST EoS) and dashed (for the FSU2R EoS) grey lines in Fig.~\ref{fig:DUrcaOnset} correspond to the central BM densities of the stars at which the DU process is activated. The points at which the vertical grey lines cross the sequence of curves in Fig.~\ref{fig:DUrcaOnset} indicate the total gravitational mass and proton fraction at the DU onset. The blue, red, and green curves indicate 2\%, 3\%, and 4\% of DM, while the grey curves show the BM stars (for details see Fig.~\ref{fig:DUrcaOnset}). For the IST EoS the enhanced cooling starts at 1.83M$_{\odot}$, 1.80M$_{\odot}$, and 1.76M$_{\odot}$ for 2\%, 3\%, and 4\% of DM, respectively. 
The small differences observed between both models are due to the denser BM core in IST compared to FSU2R. As can be seen, this phenomenon has a drastic effect on NS cooling. These results could be compared to the cooling of pure baryonic stars described by the FSU2R EoS performed in ~\cite{Negreiros:2018cho} and by the IST EoS in~\cite{Tsiopelas:2020nzm,Tsiopelas:2021ayd}.
Fig.~\ref{fig:DUrcaOnset} shows that the proton fractions of the DU process at the onset are different for the IST and FSU2R EoSs. This is due to the fact that the former takes into account the effects of excluded volume in nuclear matter, which at a given value of baryon density leads to larger Fermi momenta of nucleons compared to the point-like case of the FSU2R EoS.

\subsection{Thermal Evolution}
\label{subsec:Results_cooling}

\begin{figure}
\centering
\includegraphics[width=\columnwidth]{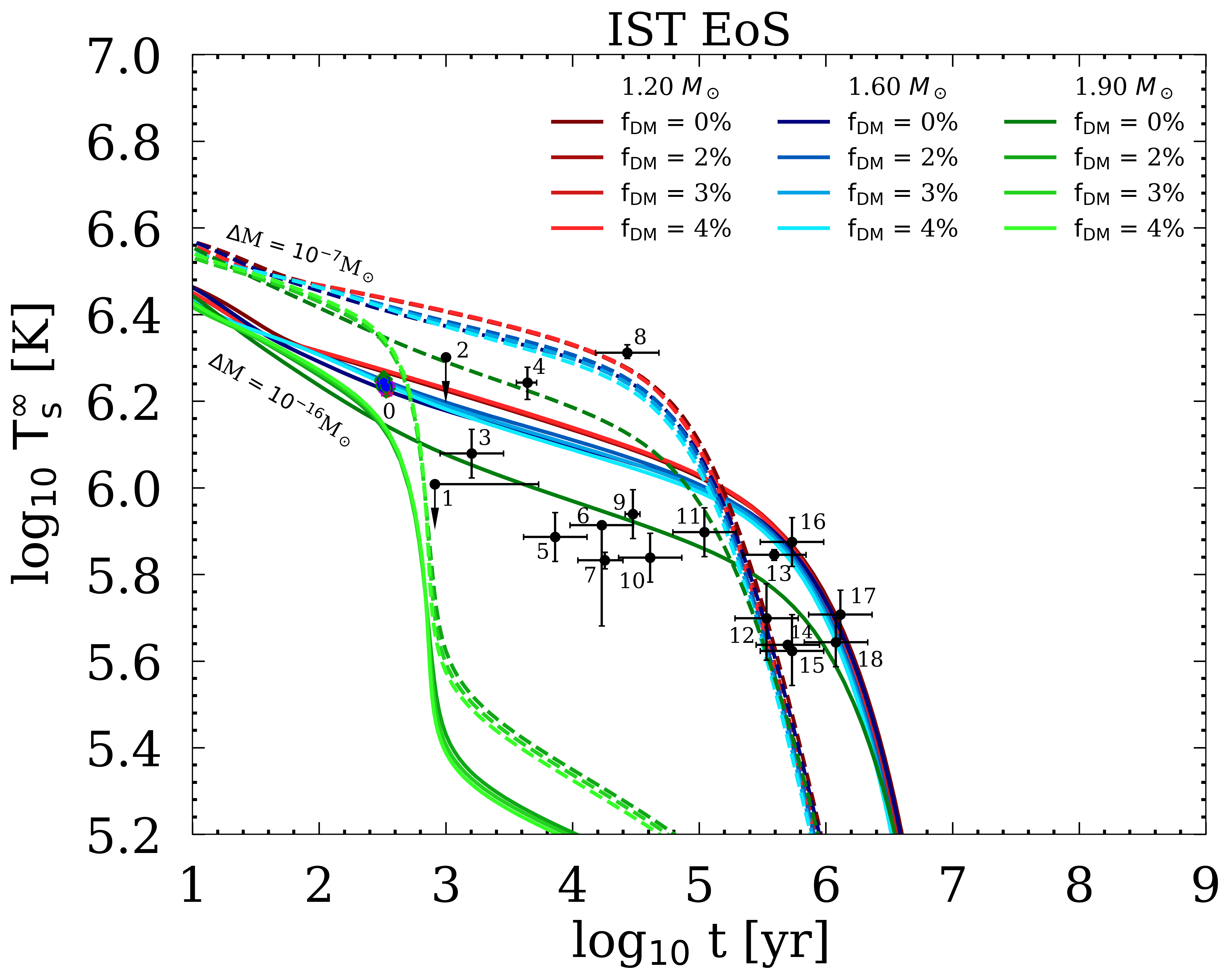}
\includegraphics[width=\columnwidth]{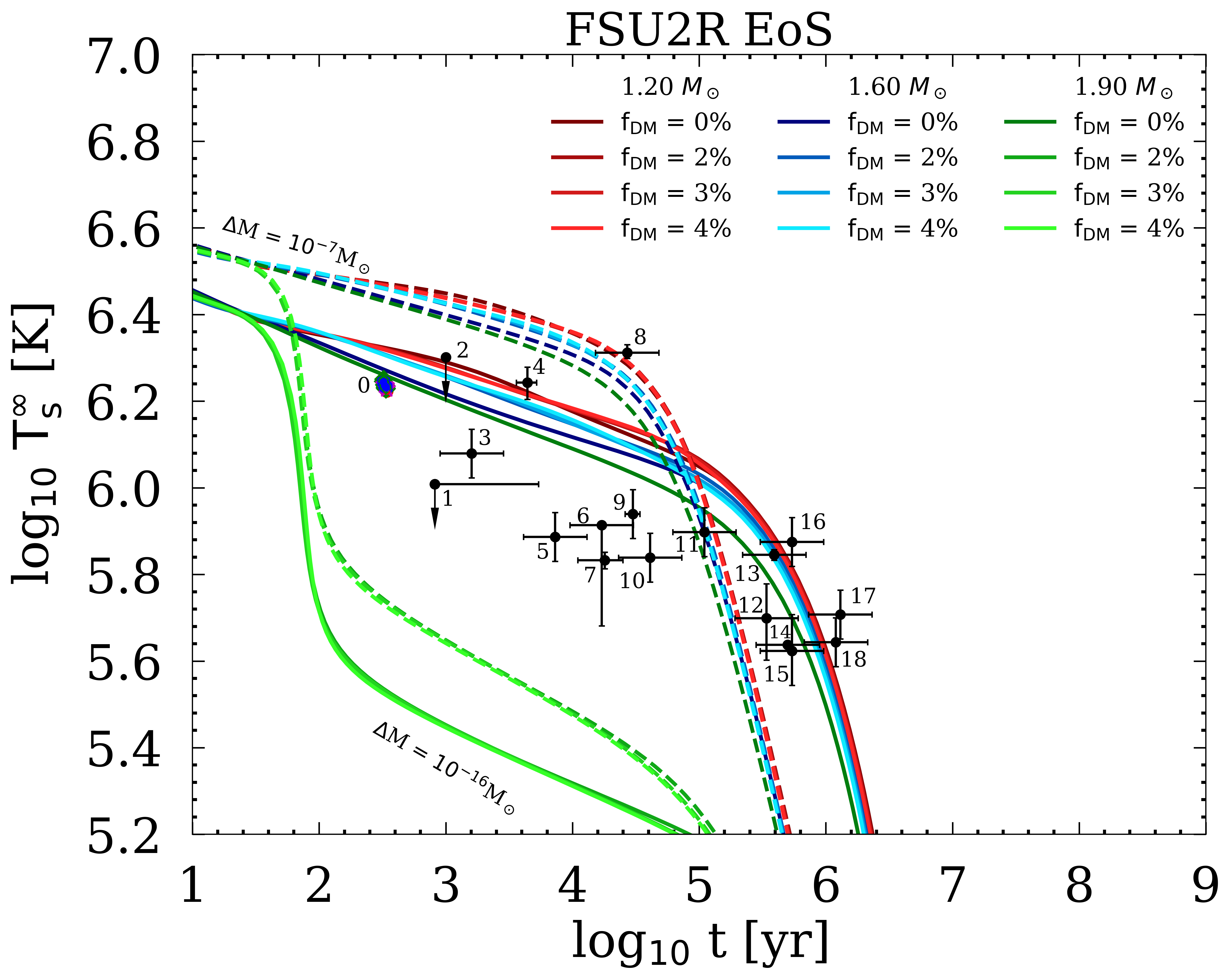}
\caption{Cooling curves for stars of different mass $M = 1.2, 1.6$ and $1.9~{\rm M}_\odot$. The calculations are performed for the IST EoS (upper panel) and FSU2R EoS (bottom panel) and the relative DM fractions of $f_{DM}=0\%, 2\%, 3\%$ and $4\%$. The considered mass of DM particles is $m_{DM} = 1$ GeV. The axes correspond to the redshifted surface temperature vs. year. The solid and dashed lines refer to heavy ($\eta=\Delta M/M = 10^{-16}$) and light-element ($\eta=\Delta M/M=10^{-7}$) envelopes, correspondingly. The effect of neutron superfluidity in the $^{1}S_0$ channel via the SFB model~\citep{Schwenk:2002fq} and proton superconductivity in the $^{1}S_0$ channel with the CCDK model~\citep{Chen:1993bam} were taken into account. The target masses refer to total gravitational masses.}
\label{fig:CoolingISTFSU2RDM}
\end{figure}

Fig.~\ref{fig:CoolingISTFSU2RDM} shows the redshifted surface temperature $T_s^\infty$ as a function of stellar age for 1.2, 1.6, and 1.9 M$_\odot$ stars. The presented cooling curves show a fit of the observational data obtained considering $^1S_0$ neutron and proton pairings, described by the SFB~\citep{Schwenk:2002fq} and CCDK~\citep{Chen:1993bam} models. The color grade in Fig.~\ref{fig:CoolingISTFSU2RDM} depicts the different DM fractions. With an increase of the DM fraction, the curves are lighter. Thus, pure BM stars for the IST EoS (upper panel) and FSU2R EoS (bottom panel) are shown in red, blue, and green for 1.2, 1.6, and 1.9${M_\odot}$ stars, respectively. As it can be seen, in both models the DU process does not operate at 1.9${M_\odot}$. However, an accumulation of DM particles with $m_{DM} = 1$ GeV of $f_{DM}\simeq0.161\%$ (IST EoS) and $f_{DM}=0.378\%$ (FSU2R EoS) triggers the previously forbidden process to operate. 

We also address the possibility of different envelope composition that affects the relation between the surface and core temperatures, and, consequently, the observed surface luminosity. Thus, the fraction of light elements is accounted for by the factor $\eta=\Delta M/M$, whereas $\Delta M$ is the mass of light elements in the upper envelope. The light-element envelope, i.e. hydrogen, helium, with $\eta=\Delta M/M= 10^{-7}$ is depicted with the dashed curves in Figs.~\ref{fig:CoolingISTFSU2RDM}-\ref{fig:CassA}, while the heavy-elements envelope, mostly carbon, with $\eta=\Delta M/M= 10^{-16}$, is shown with solid curves. 

Figs.~\ref{fig:CoolingISTFSU2RDM}-\ref{fig:CassA} depict the recently updated measurement of the HESS J1731-347 star (object number 8) which is reported to be the lightest and smallest compact object ever detected~\citep{Doroshenko2022}. Our analysis shows that the surface temperature of the HESS J1731-347 compact star can be described by the light-element envelope, which is in agreement with the results of~\cite{Sagun:2023rzp}.

\subsection{Cas A}
\label{subsec:CASA}

The compact object in the center of Cas A, being about 356 years old, is an interesting and subject-to-debate object that shows an unusually rapid cooling~\citep{Ho:2009mm,Heinke:2010cr,Shternin:2010qi,Elshamouty:2013nfa,Wijngaarden:2019tht}. The recent combined analysis of the x-ray spectra of Cas A suggests the surface temperature drop is 1.4-2.5\% over two decades of observations, the mass of the star of $M=1.55\pm$0.25 M$_{\odot}$~\citep{Shternin:2022rti}. Many models have been suggested to explain this behaviour via a strong $^3P_2$ pairing between neutrons in the core~\citep{Page:2010aw,Ho:2014pta}, rapid cooling via the DU process~\citep{Taranto:2015ubs}, medium-modified one-pion exchange in dense matter~\citep{Blaschke:2011gc}, beyond the Standard Model Physics~\citep{Hamaguchi:2018oqw}, etc.

As shown in~\cite{Tsiopelas:2020nzm, Tsiopelas:2021ayd} the IST EoS does not necessarily require the inclusion of neutron superfluidity and/or proton conductivity to explain the Cas A temperature drop. However, the obtained mass of the star fitted to the data was found to be 1.96 M$_{\odot}$~\citep{Tsiopelas:2021ayd}. In fact, in many models, the DU process is allowed only at high masses due to the existing threshold. The inclusion of DM solves this problem and allows the DU process to be activated in medium/low mass stars. 

\begin{figure}
    \centering
    \includegraphics[width=\columnwidth]{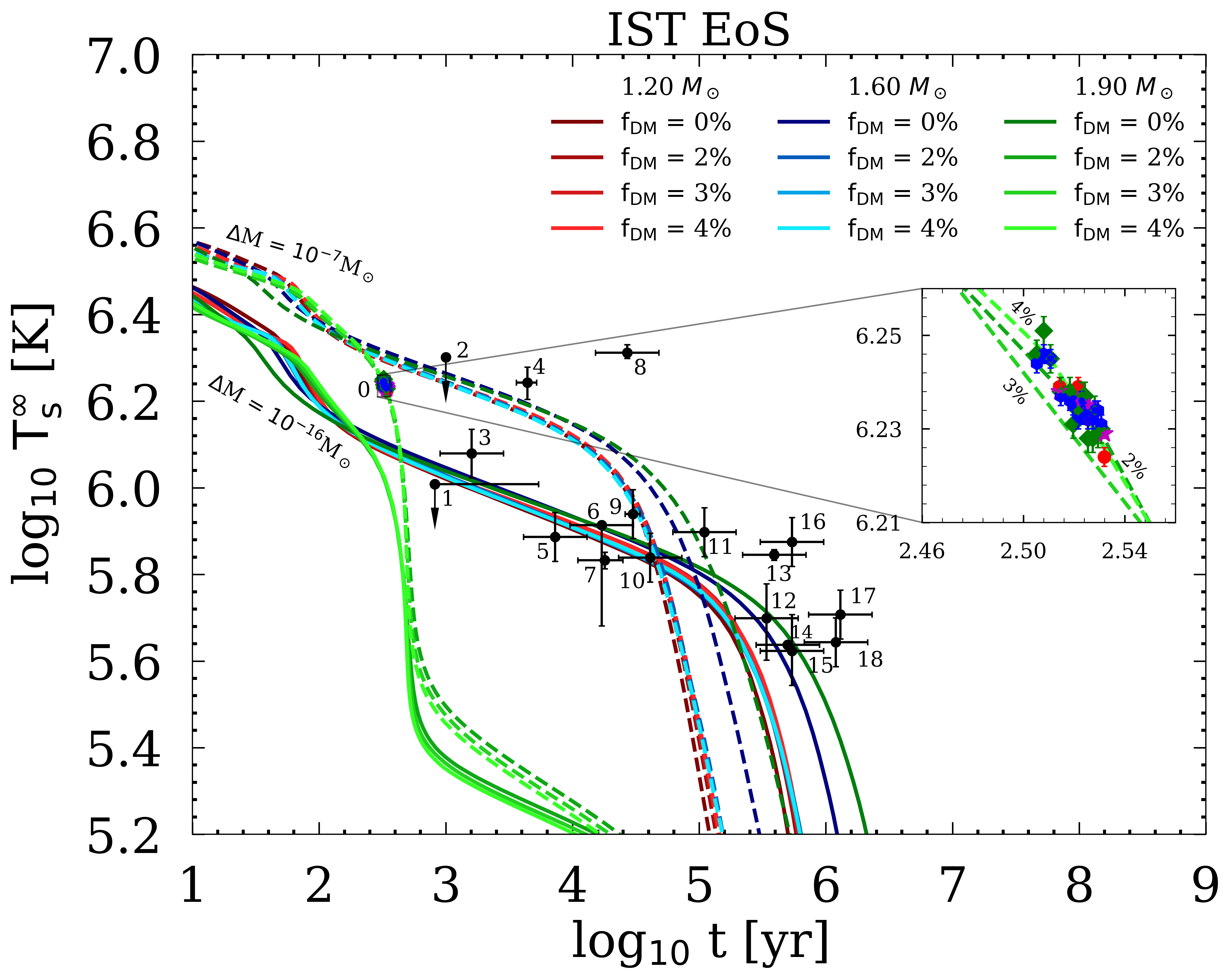}
    \includegraphics[width=\columnwidth]{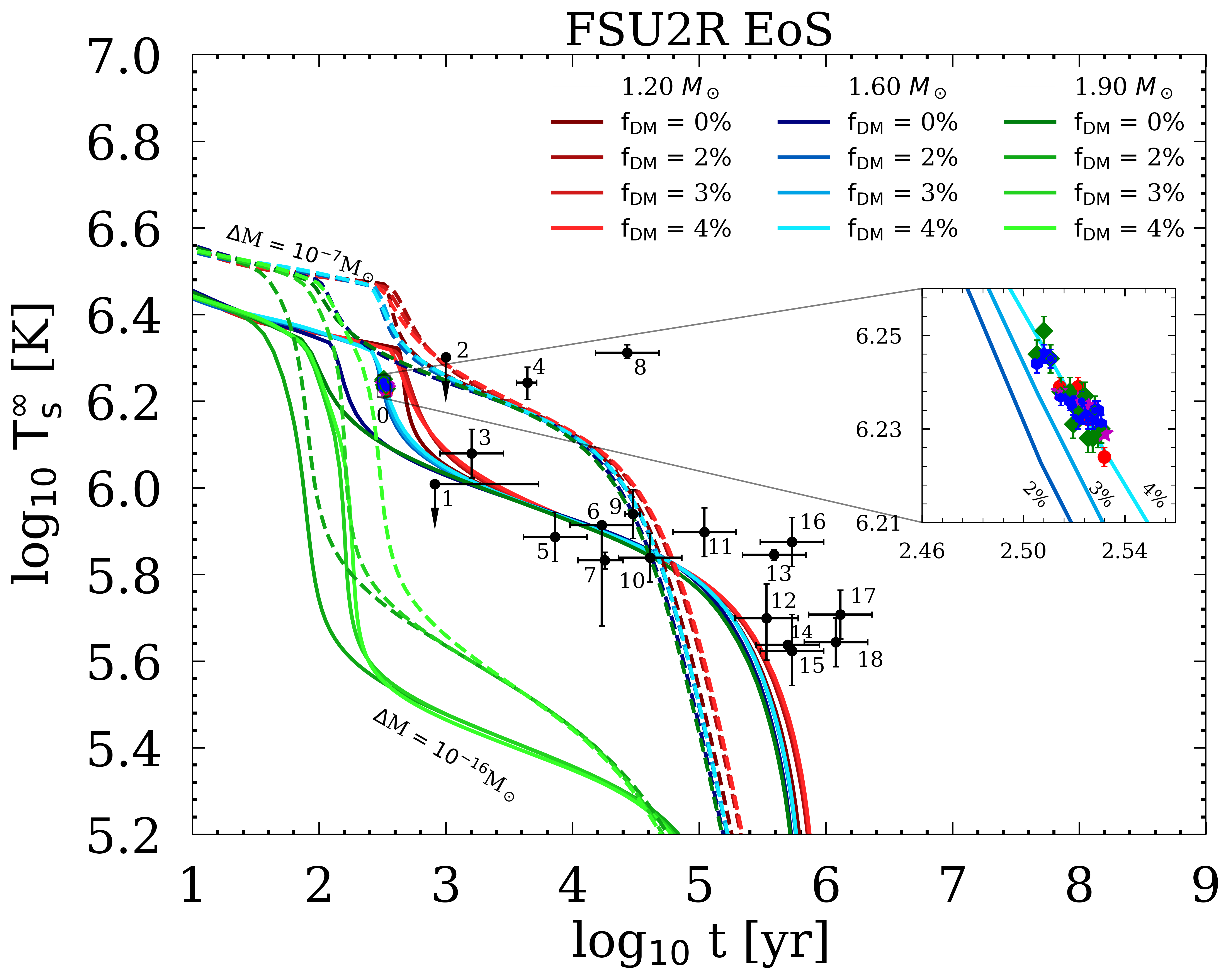}
    \caption{The same as Fig.~\ref{fig:CoolingISTFSU2RDM} supplemented with the triplet neutron pairing described by the T72 model~\citep{10.1143/PTP.48.1517}. The insets on both panels show the observational data and the best fit within the IST and FSU2R EoSs of the Cas A measured by Chandra ACIS-S in GRADED and FAINT modes. The green and blue data points correspond to the variable and fixed absorbing hydrogen column density 
    $N_{H}=1.656 \cdot 10^{22}~ cm^{-2}$ in the FAINT mode, whereas the green and blue data points depict the same data for the GRADED mode taken from~\citet{Shternin:2022rti}.}
    \label{fig:CassA}
\end{figure}

The insets on Fig.~\ref{fig:CassA} demonstrate the best fit of the Cas A observational data points. These curves were obtained by supplementing $^1S_0$ neutron and proton pairings with $^3P_2$ neutron pairing described by the T72~\citep{10.1143/PTP.48.1517} model. Thus, the upper panel of Fig.~\ref{fig:CassA} shows the cooling curves for stars of $M = 1.2, 1.6, 1.9~{\rm M}_\odot$ calculated for the IST EoS and supplemented with n $^{1}S_0$ (SFB model), p $^{1}S_0$ (CCDK model), n $^{3}P_2$ pairing (T72 model) with the maximum critical temperature $T_{c} = 6.596 \cdot 10^{8}$ K. The best agreement with the observational data is obtained for the $M =1.9~{\rm M}_\odot$ star with $f_{DM}=2-4\%$. For the FSU2R EoS, the best fit is achieved for the same combination of the pairing gaps as for the IST EoS with a slightly higher value of the maximum critical temperature for the n $^{3}P_2$ pairing,  $T_{c} = 7.105 \cdot 10^{8}$ K. The light blue curve on the inset on the lower panel of Fig.~\ref{fig:CassA} shows the $M =1.6~{\rm M}_\odot$ star with $f_{DM}=4\%$ and light-elements envelope that provides the best fit to the Cas A data and agrees with its estimated mass.

\subsection{Scan over DM parameters}
\label{subsec:Results_scan}

\begin{figure}
    \centering
    \includegraphics[width=\columnwidth]{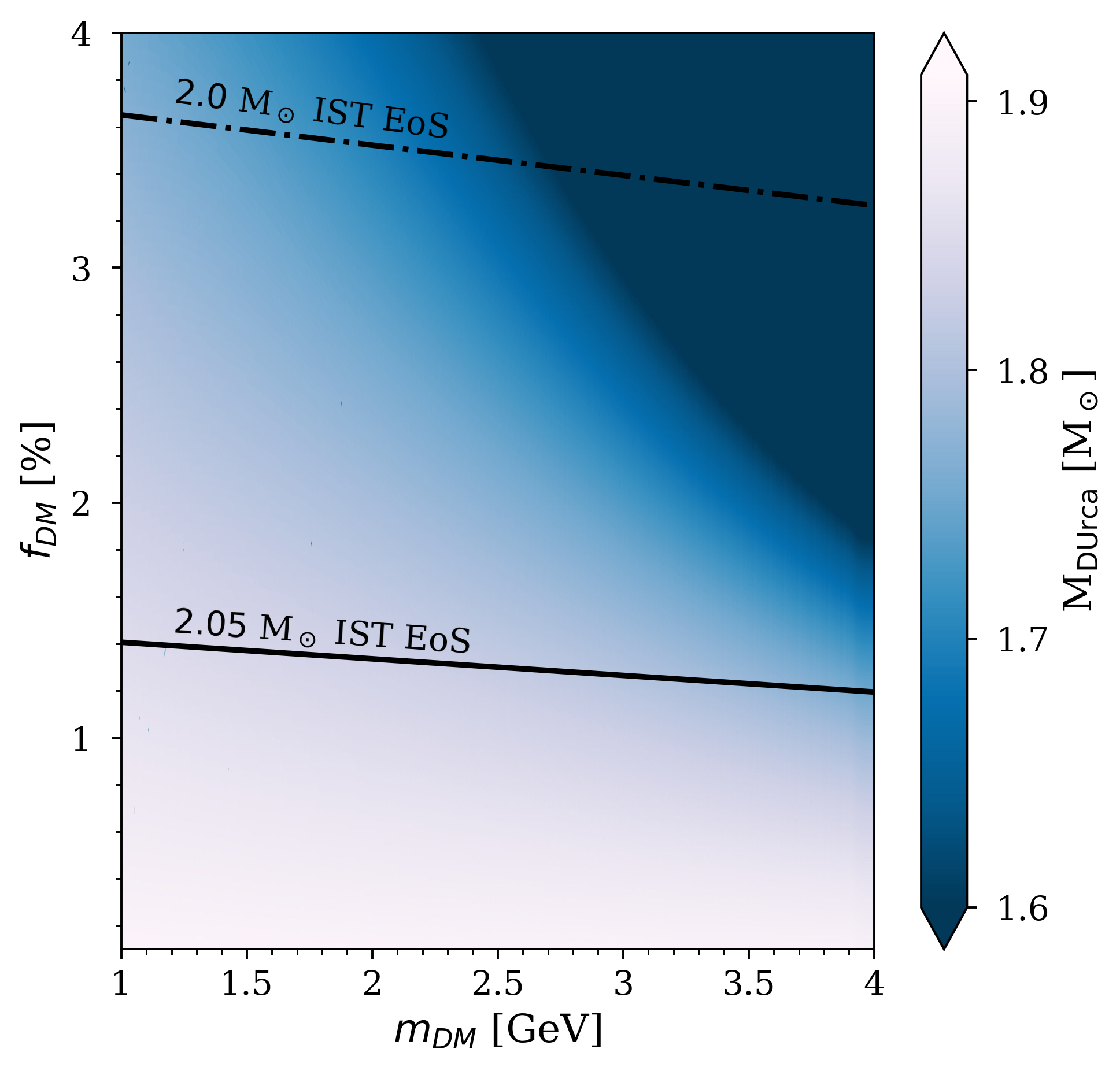}\\
    \includegraphics[width=\columnwidth]{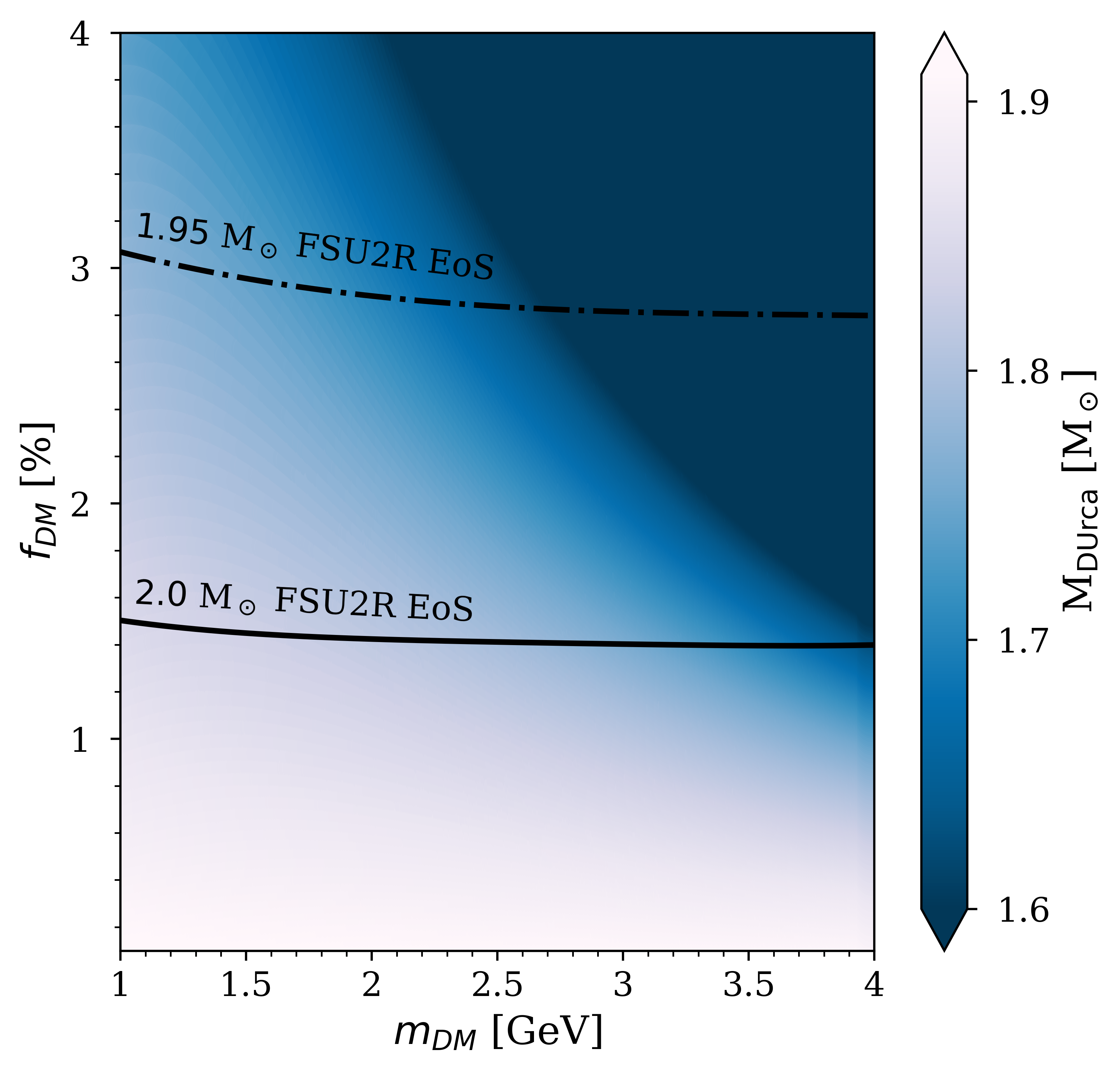}
    \caption{Scan over the particle's mass, $m_{DM}$, and fraction, $f_{DM}$, of DM-admixed stars, shown for the IST EoS (upper panel)
    and FSU2R EoS (bottom panel). The color shows the star's total gravitational mass at which the DU process starts to operate. 
    Dash-dotted and solid black curves are the contour lines showing the maximum gravitational mass obtainable for the two models.}
    \label{fig:DMscan}
\end{figure}

Fig.~\ref{fig:DUrcaOnset} shows how an increase in DM fraction for $m_{DM}=$1 GeV particles leads to a decrease in the star's mass. As it was discussed by~\citet{Giangrandi:2022wht}, the DM particle's mass and relative fraction inside the star have a comparable impact. Thus, similar DM-admixed configurations could be obtained by increasing the particle mass for the fixed value of the fraction, or vice versa. To see this effect on the DU onset and total gravitation mass for the IST EoS (upper panel) and FSU2R EoS (bottom panel) we perform scans shown in Fig.~\ref{fig:DMscan}. The color maps denote the total gravitational mass of the stars at which the DU process is kinematically allowed. The black dash-dotted and solid curves depict the attained maximum total gravitational mass for these stars.  From  Fig.~\ref{fig:DMscan} we can see that, while the onset of the DU process for the pure BM stars occurs at 1.91 M$_{\odot}$ (IST EoS) and 1.92 M$_{\odot}$ (FSU2R EoS), it could decrease below 1.6 M$_{\odot}$ for $m_{DM}\geq3$ GeV and/or DM fraction $f_{DM}\geq2$ \%. Therefore, to see a significant effect on NS cooling the fraction of heavy DM does not need to be high.

If we consider an average estimated mass of Cas A central object as 1.6 M$_{\odot}$~\citep{Shternin:2022rti} an accumulation of DM in the range of fractions and masses is shown in the top right corners on both panels of Fig.~\ref{fig:DMscan}.

\section{Conclusions}
\label{sec:Conclusion}

In this study, we focused on the effects of asymmetric fermionic DM on the NS thermal evolution. Despite asymmetric DM that interacts with BM only gravitationally contributes neither to neutrino and photon emission directly nor deposits energy to the system, it alters the thermal evolution of NSs indirectly. The calculations were performed under the assumption of a negligibly low DM accretion rate during the NS evolution, or, equivalently, a fixed DM fraction after the NS formation. In this scenario, there is no effect of dark kinetic heating on the NS's thermal evolution.
We demonstrate that an accumulated DM pulls inwards BM from the outer layers, significantly increasing the central density, hence modifying the BM distribution. Consequently, the onset of the DU process is triggered at lower NS masses, leading to a highly efficient and rapid cooling, which is substantially different from the case when it is forbidden. At the same time, the proton fraction corresponding to the DU onset remains the same, as for the pure BM star with the same central BM density. We show that despite the DU process is kinematically allowed only at 1.91 M$_{\odot}$ for the IST EoS and 1.92 M$_{\odot}$ for the FSU2R EoS, an accumulation of DM particles with $m_{DM} = 1$ GeV of $f_{DM}\simeq0.161\%$ (IST EoS) and $f_{DM}=0.378\%$ (FSU2R EoS) triggers the previously forbidden process. An increase of the DM particle's mass $m_{DM}\geq3$ GeV and/or DM fraction $f_{DM}\geq2$ \% shifts the DU onset even below 1.6 M$_{\odot}$. 

The effect of DM is also illustrated for the thermal evolution of the compact object in the center of Cas A. We find the best fit of the surface temperature drop of Cas A with the FSU2R EoS supplemented with neutron and proton singlet pairing, and triplet neutron pairing for the $M =1.6~{\rm M}_\odot$ star with a DM fraction of $4\%$. This result agrees very well with the estimated mass of the object of $M=1.55\pm$0.25 M$_{\odot}$~\citep{Shternin:2022rti}. 

An additional effect of DM is related to the pull of BM inward, creating a more compact core and reduction of the baryonic radius. Thus, the total surface of the star is reduced leading to a lower photon luminosity. This effect is clearly visible at the photon-dominated stage when the neutrino emission takes a subdominant role. 

We showed that either an increase of the DM fraction or the particle's mass causes a shift of the DU onset toward lower star gravitational masses. This effect could be a smoking gun signature of the presence of DM in compact stars. Thus, low/middle mass NSs that are not expected to have an operating DU process in fact might have it due to the presence of DM. As a result, stars of the same mass will show a different cooling pattern depending on the DM fraction accrued in their interior. This effect could be also used to map the DM distribution. In this context, present, e.g. XMM-Newton, NICER~\citep{Riley:2021pdl, Miller:2021qha}, and future, e.g. ATHENA~\citep{Cassano:2018zwm}, eXTP~\citep{eXTP:2018kws}, and STROBE-X~\citep{STROBE-XScienceWorkingGroup:2019cyd}, x-ray observational programs look very promising as they expect to increase the number of mass, radius, and surface temperature determinations. As we expect a higher DM fraction inside compact stars toward the Galactic Center, their thermal evolution could exhibit a distinct feature from the stars in the solar vicinity. 
   
\section*{Acknowledgements}
The work is supported by the FCT – Fundação para a Ciência e a Tecnologia, within the project No. EXPL/FIS-AST/0735/2021 with DOI identifier 10.54499/EXPL/FIS-AST/0735/2021. A.Á., E.G., V.S., and C.P. acknowledge the support from FCT within the projects UIDP/\-04564/\-2020 and UIDB/\-04564/\-2020, respectively, with DOI identifiers 10.54499/UIDP/04564/2020 and 10.54499/UIDB/04564/2020. E.G. also acknowledges the support from Project No. PRT/BD/152267/2021. C.P. is also supported by project No. 2022.06460.PTDC with the DOI identifier 10.54499/2022.06460.PTDC. The work of O.I. was supported by the program Excellence Initiative--Research University of the University of Wrocław of the Ministry of Education and Science.

\section*{Data availability}
The data underlying this article will be shared on reasonable request to the corresponding author.



\bibliographystyle{mnras}
\bibliography{bibliography}

% Don't change these lines
\bsp	% typesetting comment
\label{lastpage}
\end{document}